# Do Mutual Funds Make Active and Skilled Liquidity Choices in Portfolio Management? Evidence from India

Agarwal, Pankaj K[a*], Pradhan, H K[b], Saxena, Konark[c]

[a]*Finance, XLRI Delhi NCR, Jhajjar, India*

[b]*Finance & Economics, XLRI Xavier School of Management, River Meet Road, Jamshedpur, India*

[c]*Finance, ESCP Business School, Paris, France*

Correspondence Details:

[a]Finance, XLRI Delhi NCR, Dadri Toe, Untloada, Jhajjar, 124103, Haryana, India

Email: pankaj@xlri.ac.in

[b]Finance & Economics, XLRI Xavier School of Management, River Meet Road, Jamshedpur, 831001, Jharkhand, India

Email: pradhan@xlri.ac.in

[c]Finance, ESCP Business School, 79, avenue de la Republique, Paris, 75011, Ile-de-France, France

Email: ksaxena@escp.eu

# Do Mutual Funds Make Active and Skilled Liquidity Choices in Portfolio Management? Evidence from India

Abstract

This study examines active liquidity management by Indian open-ended equity mutual funds. We find that fund managers respond to inflows by increasing cash holdings, which are later used to purchase less-liquid stocks at favourable valuations. Funds with less liquid portfolios tend to maintain larger cash reserves to manage flows. Funds that make active liquidity choices yield statistically and economically significant gross and net returns. The performance differences between funds with varying activeness in altering liquidity highlight the importance of active liquidity management in markets with substantial cross-sectional liquidity differences such as India.

**Introduction**

Mutual funds globally manage trillions of dollars, navigating diverse economic environments that vary significantly across markets. In developed markets, the role of fund managers in allocating between cash and risky assets is often downplayed, as investors can easily access cash markets independently (Nanda, Narayanan and Warther, 2000). However, in less liquid and less efficient markets, effective cash management by fund managers can provide substantial value. This is because it allows managers to allocate funds to illiquid securities at more favorable prices, potentially enhancing investor returns (Dong, Feng, and Sadka, 2019). Additionally, maintaining adequate cash reserves helps prevent forced asset sales at unfavorable prices during periods of high liquidity costs, thus preserving fund performance and providing a buffer against market volatility (Chen, Goldstein, and Jiang, 2010).

The ability of mutual funds to generate alpha is influenced by the competitive environment within the active fund management industry and the amount of available liquidity pursuing such opportunities (e.g., Feldman, Saxena and Xu, 2020; Pastor and Stambaugh 2012; Pastor, Stambaugh and Taylor 2020). Limited liquidity further exacerbates decreasing returns to scale in alpha generation. As funds scale up their active positions, their trades tend to create price impacts that reduce subsequent security-level alpha (e.g., Coval and Stafford 2007; Lou 2012).

Given these dynamics and the fact that the size and competitiveness of the Indian active fund management industry are relatively smaller and less intense than in the U.S., we hypothesize that Indian active mutual funds may have greater success in using liquidity reserves to pursue alpha. Unlike the highly liquid U.S. market, liquidity in the Indian stock market is concentrated in a few dominant stocks. This presents a unique challenge for fund managers with broader mandates, as they must navigate between investing in a few liquid, low-alpha stocks or a larger number of illiquid, potentially higher-alpha opportunities. This characteristic of the Indian market offers a compelling backdrop to study how mutual funds actively balance allocations between cash, liquid stocks, and illiquid stocks in their pursuit of returns.

In this paper, we present empirical evidence on how Indian fund managers actively manage liquidity in response to investor flows and how these liquidity management strategies influence the funds' ability to generate both gross and net alpha. Our analysis evaluates the potential of active liquidity management as a tool for enhancing fund performance in markets like India.

To assess the impact of investor flows on mutual funds' liquidity management, we consider three primary responses: (i) adjusting cash holdings, (ii) liquidating a portion of the portfolio while maintaining its original composition (Popescu and Xu, 2023), and (iii) rebalancing the portfolio by selling more liquid, lower-performing securities to minimize transaction costs. While most studies on U.S. and other developed market mutual funds such as Ben- Rubi, Mugerman, and Wiener (2024), Coudert and Salakhova (2020), Jiang, Li, and Wang (2021), Lou (2012), Clarke, Cullen, and Gasbarro (2007), and Gompers and Metrick (2001) document asset sales to meet redemptions, only a few, like Chernenko and Sunderam (2016) and Popescu and Xu (2023), provide evidence of using cash reserves. We find that Indian funds dynamically adjust their cash holdings to manage both inflows and outflows. During periods of inflows, managers typically increase cash reserves and temporarily invest in liquid stocks, leading to a short-term reduction in portfolio illiquidity. This effect lasts about a quarter, after which illiquidity rises as managers gradually shift toward more illiquid, carefully selected stocks based on favorable valuations. Conversely, during periods of outflows, funds draw down cash reserves to meet redemptions, increasing portfolio illiquidity as the proportion of illiquid assets in the remaining portfolio grows.

The more illiquid the portfolio, the more pronounced the adjustments in cash holdings in response to flows. High flow volatility further complicates liquidity management by increasing uncertainty about future redemptions, prompting funds to conserve liquidity, which in turn affects asset illiquidity. After documenting that active managers dynamically manage liquidity, we analyze whether this behavior impacts their performance. We find that managers who actively manage their portfolio liquidity achieve higher returns, both before and after fees. We measure the activeness of liquidity management using the standard deviation of a fund's liquidity over time, controlling for market conditions. While a fund's liquidity level is influenced by its investment universe and benchmark, the variability in liquidity reflects the manager's active choices. This concept is akin to tracking error, where the standard deviation of returns relative to a benchmark indicates managerial activeness. Our findings show that funds with higher liquidity volatility tend to generate greater gross and net alphas. This result is robust whether analyzed through portfolio sorts or panel regression with time and fund fixed effects.

    Our findings contribute to the literature in several ways. First, we introduce a novel measure of liquidity-related activeness: the standard deviation of a fund's liquidity. Using this measure, we demonstrate that active liquidity management can enhance fund performance in the Indian active fund management industry, characterized by smaller fund sizes and a high concentration of liquidity in a few stocks. Our results suggest that the determinants of managerial skill in markets outside the U.S. may differ from those within,

relating to the heterogeneous inefficiencies that can generate alpha, such as varying levels of liquidity.

**Related Literature and Hypothesis Development**

How do the primary responses- noted above - of fund managers to investor flows affect portfolio liquidity? The first response, adjusting cash holdings, leads to changes in the relative weights of portfolio constituents, which in turn leads to a change in portfolio liquidity, defined as the weighted average illiquidity of the constituent stocks (Jiang, Li, and Wang, 2021), *ceteris paribus*. Inflows/Outflows will increase/reduce cash holdings, leading to a decrease/increase in the relative weight of constituent securities, causing an increase/decrease in portfolio liquidity. Some quick liquidity-induced trading could also occur, e.g., in the case of inflows, the fund manager may quickly buy liquid stocks momentarily to minimize the immediate impact of new cash on performance. Liquidity-induced trading is costly in the form of commissions and bid-ask spreads (Rakowski, 2010), particularly in emerging markets characterized by pervasive illiquidity. Therefore, a skilled manager will subsequently wait for the attractive (well-considered, driven by valuation) opportunities as they present themselves and shift from previously invested liquid stocks. Adjusting cash holdings could have two counterbalancing effects a) inducing higher tracking error due to associated carrying costs or cash drag (Ferson and Warther, 1996) and b) avoiding outflow-induced selling of stocks shields fund performance from price impact.

As an alternative strategy, if the fund liquidates a fractional slice of the stock portfolio to finance outflows, the portfolio illiquidity shall, in fact, fall owing to the simultaneous conservation of cash holdings and a decrease in the relative weight of the stocks in the portfolio. Consequently, the tracking error shall increase, and the fund may face a possible price impact as it sells a part of the portfolio comprising stocks with varying levels of liquidity. Also, in the presence of a liquidity premium, the fund performance may suffer as asset illiquidity falls. Therefore, using asset holdings to accommodate flows is also costly. A variation of this approach could be using a fractional slice of the *entire* portfolio, including cash, for meeting redemptions. In this case, too, the overall liquidity of the portfolio is conserved. However, the only marginal advantage of this approach is preserving the overall portfolio strategy.

It is unclear whether fund managers use portfolio- or cash holdings to manage flows, and two broad strands of literature exist. Some researchers argue that fund managers prefer to hold liquid stocks and when faced with redemptions, sell them to honor withdrawals rapidly and incur lower trading costs (Clarke, Cullen, and Gasbarro, 2007; Gompers and Metrick, 2001; Scholes, 2000; Chan and Lakonishok,1997) whereas others (Jiang, Li, and Wang, 2021; Coudert and Salakhova, 2020; Ben Raphael, 2017; Vayanos, 2004) find that the fund managers facing redemptions follow a pecking order and first liquidate more illiquid stocks to preserve liquidity, particularly during market stress.

We decide to examine the issue and hypothesize that:

If the fund adjusts cash holdings to accommodate the flows, then:

$H_1$: Cash holdings will be increasing in flows.

$H_2$: Asset illiquidity will be decreasing in flows.

Next, a fund with higher levels of asset illiquidity renders higher levels of *liquidity services* to the investors as it accepts cash (liquid) inflows, invests them in illiquid instruments, and is obligated to honour daily redemptions (Chernenko and Sundaram, 2016). Now, if the funds adjust cash holdings to manage flows, then as flows happen, the immediate effect will be a fall in asset illiquidity. However, if flows are volatile, they will be difficult to predict. Now, for example, the fund manager would rarely have a ready investment list at all times whenever unanticipated inflows occur. So, they could either continue to sit on cash or must trade in response to unanticipated investor flows, having to engage in trading to control liquidity. This liquidity-induced trading acts as a drag on performance (Alexander, Cici and Gibson, 2007) and leads to a fall in overall portfolio illiquidity. However, subsequently, when the fund manager has had time to identify more promising stocks, the portfolio is very likely to be tilted towards illiquid stocks (valuation-motivated trading). But, as the asset illiquidity increases now, the fund manager will likely use newer inflows to increase cash holdings, too, so that the liquidity transformation ability of the fund is preserved. As a result, we should see not only cash holdings rising in flows but also a higher proportion of more recent flows going towards augmenting cash holdings. Accordingly, the inverse relationship between flows and asset illiquidity should also decay in time and finally reverse.

Therefore, we hypothesize that:

H$_3$: For a given level of flows, cash holdings of funds will be increasing in asset illiquidity.

H$_4$: Increase in cash holding in flows decays in time.

*Liquidity Choices and Performance:*

Liquidity provision by mutual funds, either by higher cash holdings or by liquidity-induced trading, will negatively impact fund returns. While the impact of higher cash holdings for liquidity provision is straightforward, the impact of liquidity-induced trading is less so. Edelen (1999) and Alexander, Cici, and Gibson (2007) suggest that fund returns are negatively impacted by liquidity-motivated trading. Fulkerson and Riley (2017) estimate that one dollar of liquidity-induced trading imposes a cost of USD 0.006 on an average US fund. Now, if the fund holdings were as liquid as cash, there was no need to forgo returns by holding cash. Since they are not, cash holdings are necessary. It follows, therefore, that the higher the illiquidity of fund holdings, the higher the cash holdings, and therefore, the higher the adverse impact on returns. In addition, more liquid stocks are characterized by lower information asymmetry and are preferred by institutional investors (Diamond and Verrecchia, 1991). Therefore, conserving fund performance will necessitate active liquidity choices (liquidity activeness) between cash or stocks and between more or less liquid stocks. However, do these active liquidity choices improve fund performance? Although the presence of liquidity premium in equity returns is well documented worldwide[1], including in emerging markets (Bekaert and Harvey, 2007), to the best of our knowledge, how active management of liquidity impacts fund performance has never been. Therefore, we hypothesize that active management of liquidity is a skill that favourably impacts fund performance:

$H_5$: Fund performance will be increasing in Liquidity Activeness.

**Sample, Variables & Methodology**

*Sample*

We investigate the performance and liquidity of Indian open-ended equity mutual funds from 2011 to 2023. Our sample comprises the vanishing funds, too, and is therefore free from survivorship bias. The estimation period starts in 2011 to avoid the impact of the 2008 financial crisis on the results. We remove funds with less than 3 years of return history. Summary statistics for various metrics, including monthly portfolio holdings, fund age, expense ratios, and liquidity measures, are presented in Table 1.

*Variables*

We extract the monthly portfolio holdings, fund age (*Age*), total expense ratio (*Expense Ratio*), portfolio turnover (*Turnover*), Total Net Assets (*TNA*), and daily Net Asset Values (NAVs) from the ACE MF database. The variable *Size* is computed as a natural log of monthly TNAs. We calculate flows for fund *i* in month *t* as the percentage

---

[1] For example, see Brennan and Subrahmanyam (1996), Amihud (2002), and Hasbrouck (2009).

For an excellent review, see Amihud et al. (2015)

growth of new assets, assuming that all flows take place at the end of the month. The monthly flows are computed as:

$$Flows_{i,t} = \frac{TNA_{i,t} - TNA_{i,t-1}(1+R_{i,t})}{TNA_{i,t-1}} \qquad (1)$$

Where $R_{i,t}$ is the total return of fund i in month t.

Scheme cash holdings $Cash_{i,t}$ is computed as the sum of the percentage monthly holdings of the scheme in a) Cash & Cash Holdings and Net Assets, b) Certificate of Deposit, c) Treasury Bills, d) Commercial Paper, and e) Bills Rediscounting.

For measuring performance, we raw returns (*Ret*) and three risk-adjusted measures CAPM alpha ($α_1$), Fama-French three-factor alpha ($α_3$), and Fama-French-Momentum four-factor alpha ($α_4$) computed from rolling three-year regression of fund returns on Fama-French and Momentum factors for India obtained from IIMA data library[2].

*Liquidity Measures*

We use two proxies to measure liquidity: Amihud's (2002) illiquidity and Pastor & Stambaugh's (2003) gamma. These measures have been extensively used in the literature on the liquidity of mutual funds (Sehrish et al., 2024; Popescu and Xu, 2023; Vidal-Garcia, Vidal, and Nguyen, 2016; Simutin, 2014; Chernenko and Sundaram, 2010, etc.). The Amihud measure is defined as:

---

[2] https://faculty.iima.ac.in/iffm/Indian-Fama-French-Momentum/ accessed on 26 March 2024.

$$Illiq_{i,t} = \frac{1}{D_{it}} \sum_{d=1}^{D_{it}} \frac{|R_{i,d,t}|}{DVol_{i,d,t}} \qquad (2)$$

Where, $Illiq_{i,t}$ is Amihud illiquidity ratio of stock $i$ in the period $t$, $D_{it}$ is the number of trading days in period $t$ for stock $i$, $R_{i,d,t}$ is the return of stock $i$ on day $d$ in the period $t$, and $DVol_{i,d,t}$ is the dollar volume of stock $i$ on day $d$ in the period $t$.

The Pastor and Stambaugh's (2003) Gamma is based on the resilience idea and measures the speed with which prices need to recover from a previous day's order flow shock. The computation is as follows:

$$R^e_{i,d+1,t} = \theta_{i,t} + \varphi_{i,t} R_{i,d,t} + \gamma_{i,t}\, sign\left(R^e_{i,d,t}\right). DVol_{i,d,t} + \varepsilon_{i,t} \qquad (3)$$

Where, $\gamma_{it}$ is the liquidity measure Gamma and $R^e_{i,d,t}$ is computed as $R_{i,d,t} - R_{m,d,t}$, where $R_{m,d,t}$ is the return on the value-weighted market index return on day d in month t.

We primarily focus on Amihud's measure. Goyenko, Holden, and Trzcinka (2009) demonstrate that illiquidity measures like Amihud's, computed from daily data, accurately represent liquidity benchmarks compared to high-frequency counterparts. They also find Amihud's measure superior to most available measures and resilient to regime changes, such as shifts in minimum tick sizes to decimals.

The daily closing prices and traded volumes (INR) of all the stocks held by the sample mutual fund schemes in the study period are used to compute their liquidity measures. Fund illiquidity ($Illiq_{i,t}$) is computed as the value-weighted average of Amihud's and Pastor & Stambaugh's measures separately for each constituent stock.

Table 1: Summary Statistics

| Variables | Mean | Median | SD | Q25 | Q75 |
|---|---|---|---|---|---|
| *Cash (%)* | 3.11 | 2.19 | 3.56 | 0.00 | 4.64 |
| *Illiq(Amihud) x 100* | 1.01 | 0.68 | 1.04 | 0.39 | 1.26 |
| *Illiq(PS)x 100* | 0.63 | 0.18 | 1.79 | 0.06 | 0.56 |
| *Flows (%)* | 0.70 | 0.01 | 3.65 | -1.04 | 1.64 |
| *Turnover (%)* | 75.11 | 56.00 | 78.64 | 31.85 | 90.00 |
| *Expense (%)* | 2.34 | 2.38 | 0.35 | 2.09 | 2.58 |
| *Age (Months)* | 124.40 | 121.80 | 88.54 | 50.73 | 184.73 |
| *TNA (INR Billions)* | 290.61 | 67.07 | 554.58 | 15.80 | 292.67 |
| *Ret (%)* | 1.46 | 1.48 | 4.18 | -1.49 | 4.39 |
| *$\alpha_1$ (%)* | 0.07 | 0.06 | 0.50 | -0.23 | 0.34 |
| *$\alpha_3$ (%)* | 0.13 | 0.13 | 0.51 | -0.15 | 0.40 |
| *$\alpha_4$ (%)* | 0.17 | 0.14 | 0.47 | -0.10 | 0.40 |

Table-1 presents the average fund characteristics of the entire sample from 2011-2023. The columns present values of pooled mean, median, standard deviation, 25$^{th}$ and 75$^{th}$ percentile values respectively. *Illiq* is the monthly portfolio Amihud illiquidity score, computed by taking a value-weighted average of individual illiquidity scores of asset holdings obtained from $Illiq(Amihiud)_{i,t} = \frac{1}{D_{i,t}} \sum_{d=1}^{D_{i,t}} \frac{|R_{i,dt,}|}{DVol_{i,d,t}}$ where, $D_{i,t}$ = The number of trading days in period t for stock i, $R_{i,d,t}$ = The return of stock i on day d in the period t, and $DVol_{i,d,t}$ = The INR volume of stock i on day d in the period t. *Illiq(PS)* is the monthly illiquidity measure (gamma) of Pastor & Stambaugh (2003) computed by taking a value-weighted average of individual Gamma values of asset holdings obtained from $R_{i,d+1,t}^e = \theta_{i,t} + \varphi_{i,t} R_{i,d,t} + \gamma_{i,t} sign(R_{i,d,t}^e) \cdot v_{i,d,t} + \varepsilon_{i,t}$ where, $\gamma_{i,t}$ = Gamma, $R_{i,d,t}^e = R_{i,d,t} - R_{m,d,t}$, where $R_{m,d,t}$ is the return on the value-weighted market index return on day d in month t, and $v_{i,d,t}$ = The INR volume for stock i on day d in month t. . *Flows* is monthly percentage net growth in total assets over the previous month's total assets adjusted for the month's return and computed as $Flows_{i,t} = \frac{TNA_{i,t} - TNA_{i,t-1}(1+R_{i,t})}{TNA_{i,t-1}}$, where $R_{i,t}$ is the total return of fund i in month t. and TNA is the asset under management of the funds in INR Billions. Turnover is month-wise annual percentage change in the asset holdings. *Expense* is the month-wise fund expenses as a percentage of total assets. *Age* is the age of fund in months. *Ret* is the monthly percentage change in the Net asset Values. $\alpha_1$, $\alpha_3$ and $\alpha_4$ are monthly CAPM, Fama-French-three factor and Fama-French-Momentum four-factor alphas of funds.

## *Methodology*

To test our predictions, first, we show that changes in cash holdings are positively associated with net flows (H$_1$). We use the following baseline specifications in regressions in the spirit of Chernenko and Sunderam (2016). All the specifications are estimated with pooled fixed effects on months and funds and by clustering standard errors on months.

$$\Delta Cash_{i,t-6\to t} = \beta_0 Flows_t + \beta_1 Flows_{t-1} + ---+ \beta_5 Flows_{t-5} + \sum_{k=1}^{l} \gamma_k Fund_k + \sum_{m=1}^{n} \gamma_m Month_m + \varepsilon_{i,j,t} \quad (4)$$

Here, $\Delta Cash_{i,t-6\to t}$ is the change in cash and cash equivalents over last six months as a proportion of Total Net Assets six months ago. This variable captures the absolute change in cash and cash equivalents over the last six months.

In addition to absolute changes, we also examine whether changes in the *proportion* of cash holdings over Total Net Assets are also related to flows. As the portfolio size changes, examining changes in the proportion of cash holdings will shed additional light on how flows are linked to the shifting weight of cash holdings. We substitute the variable $\Delta Cash_{i,t-6\to t}$ with the following and retain the rest of the specification in equation (4):

$$\Delta \left(\frac{Cash}{TNA}\right)_{i,t} = \left(\frac{Cash}{TNA}\right)_{i,t} - \left(\frac{Cash}{TNA}\right)_{i,t-6} \quad (5)$$

Further, we employ the following specification to test how flows are associated with asset illiquidity (H₂).

$$\Delta Illiq_{i,t-3\to t} = \beta_0 Flows_t + \beta_1 Flows_{t-1} + ---+ \beta_5 Flows_{t-5} + \sum_{k=1}^{l} \gamma_k Fund_k + \sum_{m=1}^{n} \gamma_m Month_m + \varepsilon_{i,j,t} \quad (6)$$

Additionally, to examine if the funds having higher asset illiquidity are likely to witness larger changes in the cash holdings in flows, we test the following specification (H₃ and H₄):

$$\Delta Cash_{i,t-6\to t} = \beta_0 Flows_{i,t-3\to t} + \beta_1 Flows_{i,t-3\to t} x Illiq\_Ami_{t-6} + \beta_2 Flows_{i,t-6\to t-3} +$$

$$\beta_3 Flows_{i,t-6\to t-3} x Illiq\_Ami_{t-6} + \beta_4 Illiq\_Ami_{t-6} + \sum_{k=1}^{l} \gamma_k Fund_k +$$

$$\sum_{m=1}^{n} \gamma_m Month_m + \varepsilon_{i,j,t} \qquad (7)$$

In this specification, quarterly flows have been employed for compactness. The dependent variable is the change in cash holdings as a proportion of TNA two quarters ago. The interaction terms between Flows and illiquidity capture how flows in the last two quarters affect cash holdings, given illiquidity two quarters ago. The illiquidity variable is standardized so that its coefficients can be interpreted as the effect of a one-standard-deviation change in asset illiquidity.

**Empirical Results**

*Fund Flows and Cash Choices*

Table 2, Column 1, presents the results for the effect of flows on mutual fund cash holdings. We find that every INR 100 of inflows (outflows) increases (decreases) the contemporaneous cash and cash equivalents by INR 32 ($\beta_0$ = 0.32, $t$ = 8.42). The positive sign of $\beta_0$ confirms our prediction that cash holdings increase with inflows. This effect is economically significant. The remaining INR 68 is likely used for readily trading the underlying stocks. Additionally, the effect of past flows on cash holdings also remains positive for up to five months.

In Panel A, Column 2 of Table 2, we replace changes in the levels of cash holdings with changes in the proportion of cash to Total Net Assets (TNA) in our specification (Equation 1, Notes to Table 2).

Table 2: Fund Flows and Cash Choices

| | Panel A | | | Panel B | |
|---|---|---|---|---|---|
| | $\Delta Cash/TNA$ (1) | $\Delta(Cash/TNA)$ (2) | $\Delta Illiq$ (3) | | $\Delta Cash/TNA$ (4) |
| $Flows_{i,t}$ | 0.32*** | 0.13 *** | -0.007 ** | $Flows_{i,t-3 \to t}$ | 0.16*** |
| | (8.42) | (4.83) | (2.18) | | (4.16) |
| $Flows_{i,t-1}$ | 0.24 *** | 0.01 | -0.003 | $Flows_{i,t-6 \to t-3}$ | 0.05* |
| | (4.07) | (0.88) | (1.56) | | (1.79) |
| $Flows_{i,t-2}$ | 0.29 *** | 0.03 ** | -0.001 | $Illiq_{i,t-6}$ | 0.01 ** |
| | (3.16) | (2.07) | (0.55) | | (2.57) |
| $Flows_{i,t-3}$ | 0.15* | 0.00 | 0.007 *** | $Flows_{i,t-3 \to t} xIlliq_{i,t-6}$ | 0.13 ** |
| | (1.78) | (0.03) | (3.54) | | (2.38) |
| $Flows_{i,t-4}$ | 0.01 | -0.03 ** | - | $Flows_{i,t-6 \to t-3} xIlliq_{i,t-6}$ | -0.04 ** |
| | (0.15) | (2.02) | | | (2.06) |
| $Flows_{i,t-5}$ | -0.19 ** | -0.09 *** | - | | |
| | (2.5) | (5.14) | | | |
| Month Fixed Effects | Yes | Yes | Yes | | Yes |
| Fund Fixed Effects | Yes | Yes | Yes | | Yes |
| SE Clustered on | Months | Months | Months | | Months |
| Num. obs. | 26469 | 27117 | 27373 | | 6818 |
| Adj. $R^2$ | 0.15 | 0.06 | 0.33 | | 0.20 |

Table 2 Panel A presents the results of the pooled fixed effects regression results of the specification Equation (4), i.e.,

$\Delta Cash_{i,t-6 \to t} = \beta_0 Flows_t + \beta_1 Flows_{t-1} - - + \beta_5 Flows_{t-5} + \sum_{k=1}^{l} \gamma_k Fund_k + \sum_{m=1}^{n} \gamma_m Month_m + \varepsilon_{i,j,t}$ ------(4)

where, $\Delta Cash_{i,t-6 \to t}$ is the change in cash and cash equivalents over the last six months as a proportion of Total Net Assets six months ago. This variable captures the absolute change in cash and cash equivalents over the previous six months. In Column (2), the dependent variable is $\Delta \left(\frac{Cash}{TNA}\right)_{i,t} = \left(\frac{Cash}{TNA}\right)_{i,t} - \left(\frac{Cash}{TNA}\right)_{i,t-6}$ which captures the change in the proportion of cash holdings over Total Net Assets over the last six months. In Column (3), the dependent variable is the change in the Amihud illiquidity score of the fund over the previous three months and the estimates are from Equation (6).

$\Delta Illiq_{i,t-3 \to t} = \beta_0 Flows_t + \beta_1 Flows_{t-1} + - - - + \beta_5 Flows_{t-5} + \sum_{k=1}^{l} \gamma_k Fund_k + \sum_{m=1}^{n} \gamma_m Month_m + \varepsilon_{i,j,t}$ -----(6)

The independent variables are monthly net flows for the last six months as a proportion of Total Net Assets six months ago.

The Panel B Column (4) presents the results of the pooled fixed effects regression of the specification Equation (7):
$\Delta Cash_{i,t-6 \to t} = \beta_0 Flows_{i,t-3 \to t} \beta_1 Flows_{i,t-3 \to t} xIlliq_{t-6} + \beta_2 Flows_{i,t-6 \to t-3} + \beta_3 Flows_{i,t-6 \to t-3} xIlliq_{t-6} + \beta_4 Illiq_{t-6} + \sum_{k=1}^{l} \gamma_k Fund_k + \sum_{m=1}^{n} \gamma_m Month_m + \varepsilon_{i,j,t} - - - - -$ (7), where, the dependent variable is the change in level of cash holdings over previous six months as a proportion of Total Net Assets six months ago. The independent variables are flows of the last two quarters, Amihud illiquidity score of the fund two quarters ago, and their interactions. All specifications include month and fund fixed-effects and standard errors are clustered on months. All variables are winsorized at 5%. *, ** and *** indicate statistical significance at 10%, 5%, and 1%.

We find that for every 1% change in flows, the proportion of cash holdings relative to total assets changes by 0.13% ($\beta_0$ = 0.13, t = 4.83). This result is highly economically significant, given that the average proportion of cash holdings in our sample is 3.11% (Table 1). These findings suggest that flows not only affect the levels of cash but are also significantly linked to changes in fund managers' portfolio choices, which depend on the level of cash available for investment.

This is an important result as it indicates that investor flows influence the portfolio decisions of fund managers. Additionally, we observe that the relationship between flows and changes in the proportion of cash holdings persists over multiple periods, rather than being a short-term friction.

Next, the results in Table 2 Panel A, Column 3, suggest that funds use cash holdings to counterbalance illiquid investments. Changes in illiquidity are negatively associated with flows up to one quarter ($\beta_1 = -0.007$, $t = 2.18$), and the coefficients decay in magnitude over time, becoming positive thereafter ($\beta_4 = +0.007$, $t = 3.54$), exactly when the sign of the coefficient in Column 2 turns positive. Together, the results in Columns 2 and 3 strongly indicate that fund managers initially use inflows to increase cash holdings and simultaneously buy liquid, readily tradable stocks. Subsequently, they engage in valuation-driven trading, leading to purchases of relatively illiquid stocks.

Furthermore, when we interact flows with asset illiquidity, we find that a change of INR 100 in flows is associated with a change in cash holdings of INR 16 ($\beta_0 = 0.16$, t = 4.16), but the combined effect of flows and asset illiquidity is much larger (Table 2, Panel B, Column 4). For a fund with one standard deviation higher illiquidity of the underlying stock portfolio than the average fund, every INR 100 of flows is associated with a 29% change in cash holdings ($\beta_0 + \beta_1 = 0.29$). The results are both statistically and economically significant. We obtain similar (unreported) results using an alternative illiquidity measure by Pastor and Stambaugh (2003). These findings provide clear evidence that funds relying on cash holdings for managing flows maintain significantly higher (or lower) cash levels when the illiquidity of their underlying stock portfolio is higher (or lower). Similar findings are reported by Koont et al. (2022).

*Fund Flows and Liquidity Choices*

We assess whether flows influence the liquidity levels of funds. This would suggest that funds dynamically allocate between cash holdings and illiquid stocks in response to fund flows. Results in Column 3 of Table 2 confirm that fund managers' portfolio liquidity choices depend on flows. We find similar, though weaker, results for liquidity changes at quarterly and six-month intervals, as well as using alternative illiquidity measures like Pastor and Stambaugh (2003).

*Fund Liquidity Choices and Performance*

Next, we analyze whether there are significant performance differences between funds that actively adjust the liquidity levels of their portfolios compared to those that do not. To capture the activeness of liquidity changes, we define a simple and novel measure, Liquidity Activeness, as the standard deviation of portfolio illiquidity over the previous 12-month period. We begin with univariate portfolio sorts of funds to identify the preliminary relationship between liquidity activeness and performance. We use one raw and three risk-adjusted performance metrics raw returns, CAPM alpha, three-factor alpha, and four-factor alpha for both net and gross returns. At the end of each month, we sort all funds into 5 portfolios based on increasing liquidity activeness. We then compute the mean liquidity activeness values of each portfolio, along with the next month's equally-weighted mean performance metrics for each portfolio. This procedure is repeated until we exhaust the sample, ultimately producing a time series of lagged mean liquidity activeness values and mean performance. Finally, we test for differences in the performance of the top and bottom portfolios using the Newey and West (1986) t-statistic. The results are presented in Table 3. Testing differences removes the component of performance owing to market-related changes in illiquidity.

Table 3: Univariate Portfolio Sorts on Lagged Liquidity Activeness

| | Gross (Panel A) | | | | | Net (Panel B) | | | | |
|---|---|---|---|---|---|---|---|---|---|---|
| Portfolios | $\sigma(Illiq)$ | Ret | $\alpha_1$ | $\alpha_3$ | $\alpha_4$ | Portfolios | Ret | $\alpha_1$ | $\alpha_3$ | $\alpha_4$ |
| 1 | 0.00061 | 1.42 | 0.15 | 0.17 | 0.23 | 1 | 1.23 | -0.06 | -0.03 | 0.02 |
| 2 | 0.00110 | 1.53 | 0.21 | 0.25 | 0.29 | 2 | 1.34 | 0.01 | 0.05 | 0.09 |
| 3 | 0.00168 | 1.58 | 0.25 | 0.30 | 0.34 | 3 | 1.41 | 0.06 | 0.10 | 0.14 |
| 4 | 0.00264 | 1.71 | 0.31 | 0.36 | 0.39 | 4 | 1.53 | 0.11 | 0.16 | 0.19 |
| 5 | 0.00605 | 1.95 | 0.43 | 0.54 | 0.55 | 5 | 1.77 | 0.24 | 0.35 | 0.35 |
| 5-1 | | 0.53*** | 0.28 | 0.36*** | 0.32*** | 5-1 | 0.54*** | 0.29 | 0.37*** | 0.32*** |
| Newey-West t | | 2.92 | 0.88 | 5.99 | 6.11 | Newey-West t | 3.13 | 1.09 | 7.75 | 8.75 |

Table 3 presents the results of the univariate quintile portfolio sorts on lagged 12-month standard deviation of illiquidity for all months over the sample period. Panel A presents the mean standard deviation of illiquidity of five quintile portfolios, mean returns gross of expenses, and means of CAPM, three-factor and four-factor alphas computed with Gross returns. At the bottom, the divergence between mean return and alpha values of the highest and lowest quintile portfolios and associated Newey-West t-statistics are presented. Panel B reproduces these results with fund returns net of expenses. All variables are winsorized at 5%.
*, ** and *** indicate statistical significance at 10%, 5%, and 1%.

In Table 3, Panel A, we observe substantial differences in how actively funds manage liquidity. The difference in mean fund liquidity activeness between the least and most active portfolios is roughly tenfold (top quintile: 0.00061, bottom quintile: 0.00605). The standard deviation of liquidity, our Liquidity Activeness measure, serves as a proxy for active liquidity management and timing, similar to how tracking error proxies for activeness and timing in mutual funds (e.g., Cremers and Petajisto, 2009). Essentially, fund managers actively decide when to increase liquidity or invest in illiquid assets. Skilled managers should raise exposure to illiquid assets when the illiquidity risk premium is high and reduce it when the premium is low. This strategy can lead to outperformance compared to funds that maintain constant liquidity levels ($H_5$).

Our results show that the performance difference between funds based on active liquidity management is both statistically and economically significant (Table 3, Panel A). Depending on the benchmark used, these statistically significant estimates range from 4.2% per annum (CAPM alpha) to 6.6% per annum (raw returns). This result is economically significant and striking, especially compared to U.S. funds, as shown in a study by Lynch (2011), which finds that cross-sectional differences in liquidity levels do not affect performance in U.S. equity mutual funds despite the relatively broader illiquidity span of their asset holdings. Indian funds appear to be able to capture liquidity premiums despite a much narrower liquidity span of their holdings, likely due to the less efficient nature of the market, particularly for stocks outside the highly traded, low-alpha segment.

Another interesting finding is that the entire outperformance is passed on to investors, with the magnitude of outperformance being similar for both gross and net performance measurements (Table 3). Although we find (unreported) that fund expenses (fees) are positively associated with illiquidity, the similar gross and net performance measures indicate that the higher returns generated by active illiquidity choices outweigh the costs of active liquidity management and are partially passed on to investors.

Next, we conduct alternative tests to evaluate the determinants of liquidity activeness-related alpha. The results are in Table-4. We use a pooled fixed effects regression with controls for fund and time fixed effects. The specification includes month fixed effects to control for market-level factors, including changes in market illiquidity. We find that all our return measures, except CAPM alphas, are highly statistically significant and positively associated with our measure of liquidity activeness. These findings further reinforce the earlier results ($H_5$).

Table 4: Liquidity Activeness and Performance

|  | Gross (Panel A) | | | | Net (Panel B) | | | |
|---|---|---|---|---|---|---|---|---|
|  | Ret | $\alpha_1$ | $\alpha_3$ | $\alpha_4$ | Ret | $\alpha_1$ | $\alpha_3$ | $\alpha_4$ |
| $\sigma(Illiq)_{t-1}$ | 89.69*** | -15.24** | 6.37*** | 10.49** | 79.54*** | -13.88*** | 5.87*** | 9.46*** |
|  | (3.30) | (3.57) | (4.18) | (7.15) | (3.31) | (3.38) | (4.18) | (7.83) |
| Month Fixed effects | Yes | Yes | Yes | Yes | Yes | Yes | Yes | Yes |
| Fund Fixed Effects | Yes | Yes | Yes | Yes | Yes | Yes | Yes | Yes |
| SE Clustered on | Months | Months | Months | Months | Months | Months | Months | Months |
| Fund Controls | Yes | Yes | Yes | Yes | Yes | Yes | Yes | Yes |
| Num. obs. | 22264 | 22264 | 22264 | 22264 | 22264 | 22264 | 22264 | 22264 |
| Adj. $R^2$ | 0.90 | 0.60 | 0.66 | 0.61 | 0.87 | 0.62 | 0.67 | 0.62 |

Table 4 presents the results of the pooled fixed effect regression of univariate portfolio sorts on lagged standard deviation of 12-month illiquidity with fund controls Age, Size, Expenses, and Turnover. The dependent variable in Panel A is Gross Returns, CAPM alpha, three-factor alpha, and four-factor alpha, respectively. Panel B reproduces the results in Panel A by replacing gross returns with returns net of expenses. All specifications use Month and Fund fixed-effects and standard errors are clustered on Months. All variables are winsorized at 5%. *, ** and *** indicate statistical significance at 10%, 5%, and 1%.

**Discussion and Conclusion**

This paper examines how active equity fund managers in the Indian market manage liquidity in response to investor flows. Our analysis uncovers several key insights into the dynamics of active liquidity management in this distinctive context.

First, we find that Indian mutual funds primarily manage flows by adjusting cash holdings, with cash levels strategically aligned to the illiquidity of the risky assets in their portfolios. This dynamic cash allocation enables fund managers to balance liquidity needs while capturing higher returns from illiquid assets. The evidence suggests that in markets like India, where liquidity is concentrated in a few stocks, active liquidity management offers a significant advantage compared to markets where liquidity is more evenly distributed.

Our results show that this strategy of active liquidity management leads to both statistically and economically significant value creation, as reflected in the gross and net performance measures.

These findings underscore the importance of liquidity management as a core skill in active fund management, especially in markets with substantial liquidity heterogeneity. While many studies focus on the security selection abilities of fund managers, our research fills an important gap by demonstrating that liquidity management plays a critical role in driving fund performance in markets like India.

**Funding**

This research did not receive any specific grant from funding agencies in the public, commercial, or not-for-profit sectors.

**Declaration of generative AI and AI-assisted technologies in the writing process.**

During the preparation of this work the authors used chatGPT in order to copy edit the manuscript. After using this tool/service, the authors reviewed and edited the content as needed and take full responsibility for the content of the publication.

**Data Availability Statement**

Available on Request